\documentclass{vgtc}                          

\graphicspath{{figures/}{pictures/}{images/}{./}} 

\usepackage{times}                     

\usepackage{tabu}                      
\usepackage{booktabs}                  
\usepackage{lipsum}                    
\usepackage{mwe}                       

\usepackage{mathptmx}                  

\usepackage{enumitem}
\onlineid{0}

\vgtccategory{Research}

\vgtcinsertpkg

\preprinttext{To appear in the 23rd IEEE International Symposium on Mixed and Augmented Reality (ISMAR24) SafeAR workshop.}

\ieeedoi{xx.xxxx/ISMAR.xxxxxxx}

\title{AR-Facilitated Safety Inspection
and Fall Hazard Detection on Construction Sites}

\author{
    Jiazhou Liu\thanks{e-mail: joe.liu@monash.edu} %
\and Aravinda S. Rao\thanks{e-mail: aravinda.rao@unimelb.edu.au} %
\and Fucai Ke\thanks{e-mail: fucai.ke1@monash.edu} %
\and Tim Dwyer\thanks{e-mail: tim.dwyer@monash.edu} %
\and Benjamin Tag\thanks{e-mail: benjamin.tag@monash.edu} %
\and Pari Delir Haghighi\thanks{e-mail: pari.delir.haghighi@monash.edu}}
\affiliation{\scriptsize Monash University, University of Melbourne, Building 4.0 CRC}

\teaser{
 \centering
 \includegraphics[width=\linewidth]{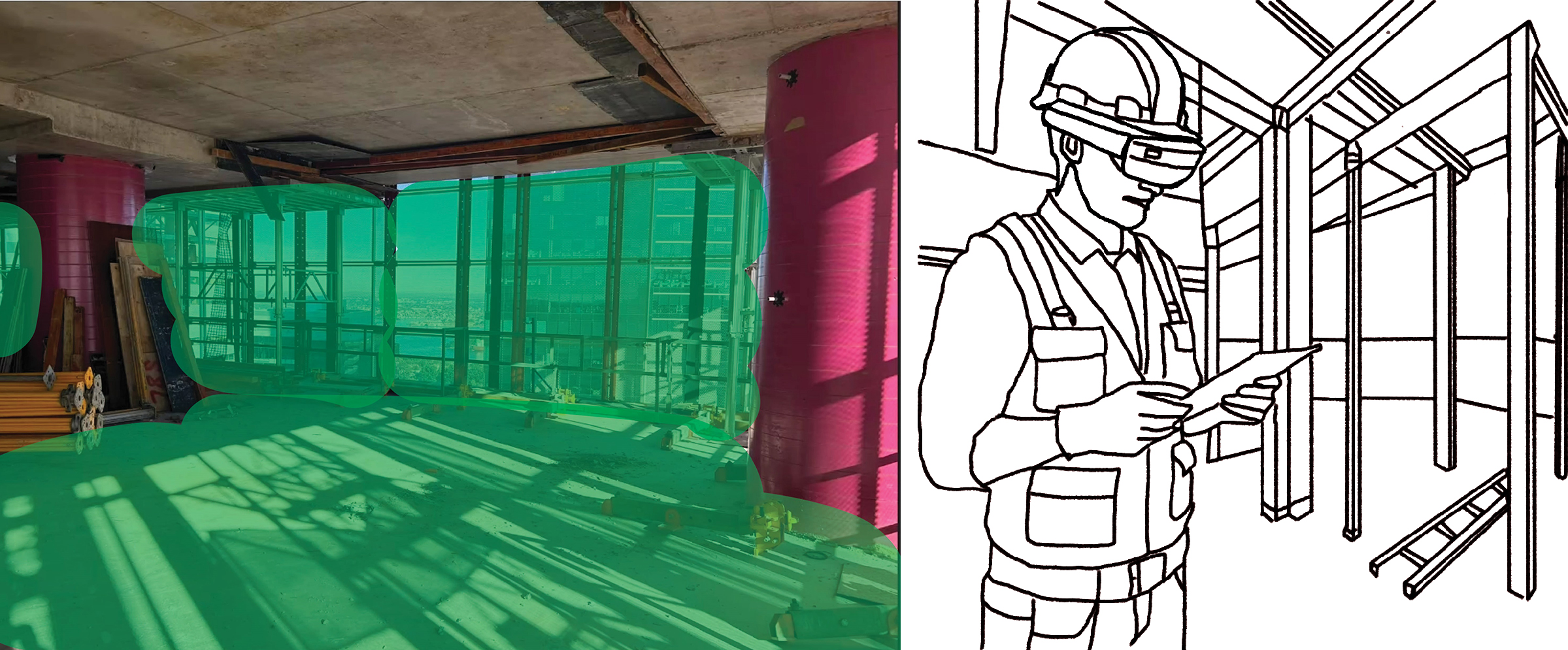}
 \caption{(left) First-person view of the proposed safety inspection solution via AR headsets. Workers virtually ``brush'' the inspected areas with coloured and situated digital overlays. (right) A use case scenario where a worker wears an AR headset to conduct a routine safety inspection.}
 \label{fig:teaser}
}

\abstract{
    Together with industry experts, we are exploring the potential of head-mounted augmented reality to facilitate safety inspections on high-rise construction sites. A particular concern in the industry is inspecting perimeter safety screens on higher levels of construction sites, intended to prevent falls of people and objects. We aim to support workers performing this inspection task by tracking which parts of the safety screens have been inspected. We use machine learning to automatically detect gaps in the perimeter screens that require closer inspection and remediation and to automate reporting. This work-in-progress paper describes the problem, our early progress, concerns around worker privacy, and the possibilities to mitigate these.
} 


\begin{document}


\firstsection{Introduction}
\maketitle
Augmented Reality (AR) is an emerging and interactive technology with the potential to significantly advance the Architecture, Engineering, and Construction (AEC) industry. The ability of AR systems to bring information ``out of the display'' and integrate it into the world around us allows digital information to be viewed within the context of objects in the surrounding environment.
Recent advances, such as low-latency, wide-field-of-view, and pass-through headsets, may help increase AR adoption across many domains.
On the other hand, as reported by industry experts, most building projects still heavily rely on manual processes during construction. For example, routine safety observations and inspections need to be performed by workers or contractors, while the inspection report will be submitted after the inspection. The accuracy and comprehensiveness of the report will be dependent on the worker's memory. 
With geo-locating, eye tracking, and built-in recording via head-mounted AR systems, inspectors can track inspected areas and virtually annotate the recorded scenes~\cite{PARK:2013:Safety}. 
Moreover, AR systems can also be integrated with Artificial Intelligence (AI) and Digital Twins to automate the safety inspection procedure and fall hazard detection on construction sites~\cite{CHEN:2024:ARCV}.
Past studies have explored the benefits of applying AR in the construction industry, such as visualising subsurface utilities~\cite{oke:2021:evaluation, oke:2022:analysis}, in-situ virtual examination of physical sites~\cite{arowoiya:2020:appraisal}, safety training including heavy equipment operation~\cite{Wu:2019:Design}. Yet, these studies also indicate that there remain challenges in applying AR in building projects, especially when it comes to data security and workers' privacy. 
Tackling the challenges in this area demands a joint effort from AEC companies, technology providers, regulators, and industry associations.

In this paper, we will first introduce an AR system for construction site safety inspection and fall hazard detection in \autoref{sec:prototype}. Then, we will discuss the potential concerns and challenges for current and future AR ecosystems for AEC projects and their possible solutions in \autoref{sec:challenge}. With this paper, we intend to make a first major step towards critically assessing the current AR system design to address potential risks holistically.

\section{AR-Facilitated Safety Inspection and Fall Hazard Detection}
\label{sec:prototype}
Falling from Height (FFH) accidents, including Fall of Person and Object, are some of the most common causes of vulnerability and fatality in high-rise building projects. In the past decades, building projects mainly relied on manual routine inspections to observe fall hazards, such as gaps in the perimeter safety screens and holes in the floors. 
With the development of advanced Computer Vision (CV) technologies and methods, as well as increased access to computing resources and publicly available datasets, we have seen a gradual increase in research in this area in recent years. 
Among the emerging technologies, camera-based CV technology has been the dominant technology for identifying FFH risks and incidents.
One of the challenges of using camera-based computer vision technology is the limited availability of public datasets necessary to train the CV models.
On the other hand, AR systems facilitate capturing video and images of the surrounding environment and automatically associate them with contextual information (e.g., geo-location data). The snapshots and recorded videos can then be used to train or test CV models.
Thus, we propose a pipeline solution for construction site safety inspection and fall hazard detection by integrating AR and advanced CV technology (Large vision model).
With the advancement of tracking ability and natural field-of-view, we use a headset-based AR, which could exploit the gaze and geo-location features to track the workers' progress.

\section{Proposed AR System}
Our proposed system targets AEC-specific requirements by providing the following features.

\noindent\textbf{Capture the dynamic construction site environment}: we will use the built-in spatial mapping features from the headset to scan the surrounding environment.

\noindent\textbf{Indicate the progress and completeness of the inspection}: we design a ``brushing'' feature to highlight areas that have been inspected (see \autoref{fig:teaser}).


\noindent\textbf{Annotations}: Users can annotate virtually or add sticky notes to a specific location in 3D space. Voice-transcribed notes can also be generated and exported together with videos or photos. 
    
\noindent\textbf{Automatic report generation system}: to \textit{summarise and produce inspection reports} based on notes and snapshots.

\noindent\textbf{Integration of Large Vision models}: (e.g., segmenting safety hazards) into our AR system to \textit{automatically detect safety issues} (e.g., gaps on the perimeter safety screens)

\section{Challenges for future AR ecosystem for AEC projects}
\label{sec:challenge}
Our vision about this AR solution and the potential future AR ecosystem used in the construction industry contains the following common tasks and contexts. 
\begin{itemize}[nosep]
    \item Natural interactions to take photos/ videos from headsets.
    \item Geolocation-based tracking via the headset's built-in sensors.
    \item Automating safety inspection and generating a report.
    \item Inspectors spend less time conducting site safety inspections.
\end{itemize}

Several challenges and concerns arise from proposing and implementing the aforementioned AR solutions.
For example, the headset continuously tracks the inspector's physical locations. Project managers can monitor the real-time tracking data, posing privacy concerns. 
Moreover, the snapshots and videos may record workers or bystanders who have not consented to being recorded.
Building unions strongly advocate for their workers' privacy to be protected.  Therefore, mechanisms must be implemented to prevent managers from using AR solutions to monitor worker activity closely.

\subsection{Key Privacy Aspects}

\noindent\textbf{Context-aware privacy-preserving mechanisms:} Protect privacy by setting the privacy preferences based on location, scene, others' presence, and hand gestures \cite{shu2018cardea}. These preferences can be dynamically adjusted in emergencies or areas of security clearances.

\noindent\textbf{Federated Learning:} Machine learning models are trained across multiple decentralized edge devices or servers while keeping the data locally on the machines. This is useful when we want to analyze data from multiple sites. 

\noindent\textbf{Differential Privacy:} Allows us to share aggregate information about a dataset while withholding information about individuals. This is especially useful when generating reports on worker productivity or safety compliance. 

\noindent\textbf{Multi-Party Computation:} Allows multiple stakeholders to analyze data jointly while keeping the source private. This is beneficial when multiple parties want to analyze construction site data while keeping the proprietary knowledge private. 

\noindent\textbf{Segmenting and Blurring Bystanders' Faces}
One possible solution to protect the privacy of bystanders is to use AI segmentation techniques to blur their faces while taking videos or snapshots.
In a system proposed by Cheng et al.~\cite{CHENG:2022:Blur} for monitoring site safety compliance, the system can automatically blur workers' faces upon saving the video frames.
With the rapid development of lightweight on-device deep learning techniques, future AR solutions could leverage these advancements to address bystander privacy concerns.

\noindent\textbf{3D Human Skeleton Features Extraction}
Another possible solution is to extract 3D human skeleton automatically features from the video recordings~\cite{DUAN:2023:Skeleton, Guo:2023:3DSkeleton, HUANG:2023:Skeleton, liu:2023:fallSkeleton}. The extracted skeleton features~\cite{Li:2023:Gesture} can then be used to train an ergonomic posture recognizer, which can identify poor postures and perform posture-based stability analysis.

\subsection{Key security aspects}

\noindent\textbf{End-to-End Encryption:} Employing robust encryption protocols for data in transit and at rest to ensure that the data remains unreadable to unauthorized parties.

\noindent\textbf{Blockchain technology:} Integrating blockchain technology \cite{mercan2021blockchain} will help create tamper-proof video access and modify audit trails, improving security and trust among stakeholders and workers.

\noindent\textbf{Access Control and Authentication:} Multi-factor authentication and role-based access control systems ensure that only authorized personnel can access sensitive data (audio, images, videos and other data). Having tiered accesses, biometric authentication, and geolocation-aware systems will be useful.

\noindent\textbf{Physical Security:} Implementing anti-tampering and anti-theft protection mechanisms to ensure monitoring system integrity. 

\noindent\textbf{Data Retention and Deletion Policies:} Implement policies and procedures on the data retention period and deleting data securely (and irreversibly) after the retention period.

\subsection{Conclusion}
This short paper proposes an AR-facilitated safety inspection and fall hazard detection solution for construction sites. We also discuss the potential security and privacy harms for both AR users and non-users and our response to those harms.
From this workshop, together with researchers and domain experts in AR and Security \& Privacy, we hope to assess and review the development of AR solutions to address potential risks holistically.

\acknowledgments{
This research is partially supported by Building 4.0 CRC. We are grateful for the industry expertise, advice and support we have received from Tim Butler and Denis Fantov from Lendlease Group as well as advice concerning computer vision models from Hamid Rezatofighi and Weiqing Wang from Monash University. We acknowledge the support of the Commonwealth of Australia through the Cooperative Research Centre Programme.
}

\bibliographystyle{abbrv-doi}

\bibliography{bibliography}
\end{document}